\documentclass[conference]{IEEEtran}
\IEEEoverridecommandlockouts
\usepackage{cite}
\usepackage{amsmath,amssymb,amsfonts}
\usepackage{algorithm}
\usepackage{algpseudocode}
\usepackage{graphicx}
\usepackage{textcomp}
\usepackage{hyperref}
\usepackage{amsmath} 
\usepackage{xcolor}
\usepackage{quantikz}
\def\BibTeX{{\rm B\kern-.05em{\sc i\kern-.025em b}\kern-.08em
    T\kern-.1667em\lower.7ex\hbox{E}\kern-.125emX}}

\renewcommand{\figureautorefname}{Figure~\negthinspace}

\begin{document}

\title{Neural Architecture Search for Quantum Autoencoders
\thanks{The views expressed in this article are those of the authors and do not represent the views of Wells Fargo. This article is for informational purposes only. Nothing contained in this article should be construed as investment advice. Wells Fargo makes no express or implied warranties and expressly disclaims all legal, tax, and accounting implications related to this article.}
}

\author{
\IEEEauthorblockN{ 
    Hibah Agha\IEEEauthorrefmark{1} \IEEEauthorrefmark{5},
    Samuel Yen-Chi Chen \IEEEauthorrefmark{2}\IEEEauthorrefmark{6},
    Huan-Hsin~Tseng\IEEEauthorrefmark{3}\IEEEauthorrefmark{7}, Shinjae~Yoo\IEEEauthorrefmark{3}\IEEEauthorrefmark{8}
}

\IEEEauthorblockA{\IEEEauthorrefmark{1}College of Engineering and Computing Sciences, New York Institute of Technology, Old Westbury, NY 11568, USA}
\IEEEauthorblockA{\IEEEauthorrefmark{2}Wells Fargo, New York, NY 10017, USA}
\IEEEauthorblockA{\IEEEauthorrefmark{3}AI Department, Brookhaven National Laboratory, Upton NY, USA}

\IEEEauthorblockA{Email:\IEEEauthorrefmark{5} \texttt{hibahswe@gmail.com}, \IEEEauthorrefmark{6} \texttt{yen-chi.chen@wellsfargo.com}, \IEEEauthorrefmark{7} \texttt{htseng@bnl.gov}, \\ \IEEEauthorrefmark{8} \texttt{sjyoo@bnl.gov}
}}

\maketitle

\begin{abstract}
In recent years, machine learning and deep learning have driven advances in domains such as image classification, speech recognition, and anomaly detection by leveraging multi-layer neural networks to model complex data. Simultaneously, quantum computing (QC) promises to address classically intractable problems via quantum parallelism, motivating research in quantum machine learning (QML). Among QML techniques, quantum autoencoders show promise for compressing high-dimensional quantum and classical data. However, designing effective quantum circuit architectures for quantum autoencoders remains challenging due to the complexity of selecting gates, arranging circuit layers, and tuning parameters.

This paper proposes a neural architecture search (NAS) framework that automates the design of quantum autoencoders using a genetic algorithm (GA). By systematically evolving variational quantum circuit (VQC) configurations, our method seeks to identify high-performing hybrid quantum-classical autoencoders for data reconstruction without becoming trapped in local minima. We demonstrate effectiveness on image datasets, highlighting the potential of quantum autoencoders for efficient feature extraction within a noise-prone, near-term quantum era. Our approach lays a foundation for broader application of genetic algorithms to quantum architecture search, aiming for a robust, automated method that can adapt to varied data and hardware constraints.
\end{abstract}


\section{Introduction}
Deep learning (DL) has become foundational to a wide range of modern applications, including computer vision, natural language processing, and scientific research. Its strength lies in its ability to uncover complex patterns in data and produce highly accurate predictions \cite{schulda2014introduction}. DL builds upon traditional machine learning (ML) by employing deep neural network architectures—comprised of multiple interconnected layers, to extract features and learn representations at various levels of abstraction. These capabilities have driven remarkable progress in tasks once deemed infeasible~\cite{janiesch2021} such as mastering complex games like Go through reinforcement learning defeating world champions \cite{silver2016mastering}, and predicting protein structures with atomic accuracy~\cite{jumper2021highly}.

Concurrently, quantum computing (QC) is rapidly evolving from theory to practice, with companies like Google, IBM, and Intel developing hardware to achieve quantum advantage \cite{yetis2021investigation}. QC offers the potential to solve problems that are intractable for classical computers by leveraging quantum superposition and entanglement, particularly in optimization, machine learning, and simulation tasks~\cite{garg2020, bayerstadler2021industry}.

\begin{figure}
    \centering
    \includegraphics[width=1\linewidth]{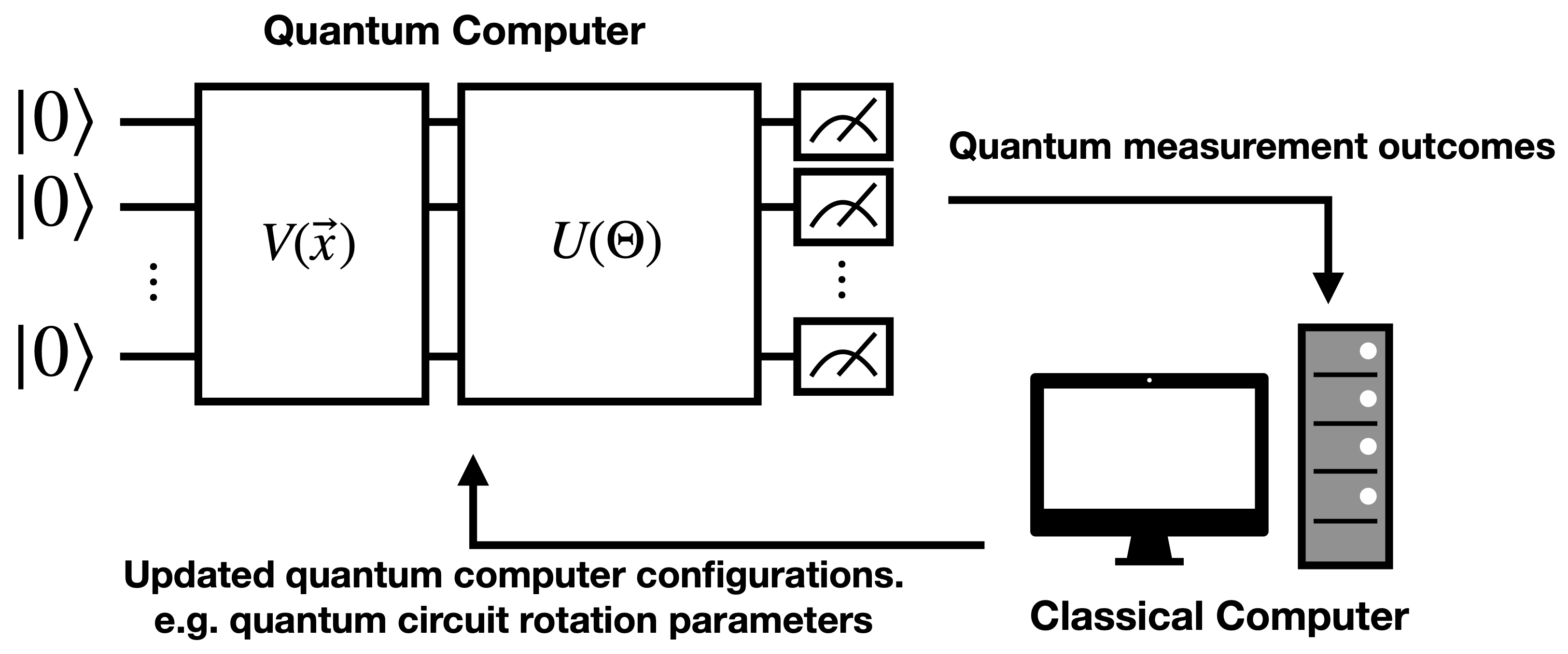}
    \caption{Hybrid Quantum-Classical Computing Paradigm.}
    \label{fig:hybrid_quantum_classical}
\end{figure}

Despite these promising capabilities, current quantum devices—referred to as Noisy Intermediate-Scale Quantum (NISQ) devices—face significant challenges. These include susceptibility to noise, decoherence, and restricted qubit counts, necessitating error correction techniques \cite{preskill2018, knill1997, campbell2017}. To navigate these limitations, hybrid quantum-classical computing has emerged as a practical and scalable approach. One of the most prominent strategies for tackling such limitations are variational quantum algorithms (VQAs) \cite{bharti2022noisy}, by leveraging quantum resources where beneficial while relying on classical systems for robustness. The general idea of the hybrid computing paradigm is illustrated in \figureautorefname{\ref{fig:hybrid_quantum_classical}}.

Building on the foundation of VQAs, the intersection of QC and ML, quantum machine learning (QML), has garnered significant research interest due to its potential to outperform classical machine learning models \cite{abbas2021power,du2020expressive,caro2022generalization}. Empirically, QML has demonstrated successful use cases in various domains such as classification \cite{mitarai2018quantum,mari2020transfer,chen2021end,qi2023qtnvqc,chen2022quantumCNN,gong2024quantum,lin2024quantumGRADCAM}, sequence learning \cite{chen2022QLSTM,chen2024QFWP,chehimi2024FedQLSTM}, natural language processing \cite{li2023pqlm,yang2022bert,di2022dawn,stein2023applying} and reinforcement learning \cite{chen2020VQDQN,chen2022variationalQRL,chen2023QLSTM_RL,skolik2022quantum,jerbi2021parametrized,yun2022quantum}. However, certain specialized applications, such as quantum autoencoders, deliver a promising, yet relatively underexplored application. 
By leveraging quantum principles, instead of classical algorithms, quantum autoencoders can accelerate feature extraction and reduce computational costs for quantum data processing \cite{du2024, romero2017, zhang2024hybrid}. These capabilities position quantum autoencoders as a powerful tool for processing large-scale datasets when working with both classical and quantum data. 

Although quantum autoencoders offer computational advantages over classical models, designing effective architectures remains challenging due to the complexity of quantum gate selection, circuit design, and parameter tuning \cite{ciliberto2018quantum}. Variational quantum circuits (VQCs), in particular, require careful configuration of gates and hyperparameters, which significantly affect performance. In this paper, we propose a quantum neural architecture search framework for quantum autoencoders that automates circuit design using genetic algorithms. Unlike differentiable methods or reinforcement learning approaches—which often struggle with local minima or require extensive hyperparameter tuning—our method robustly explores the circuit space to identify high-performing configurations. This approach results in more effective feature extraction and clustering of high-dimensional data, while leveraging the advantages of quantum computation.

\section{Related Work}
\subsection{Quantum Architecture Search}

Crafting a high-performing QML model for specific tasks necessitates deep expertise in quantum information science. To enhance QML's applicability, considerable efforts have been directed toward automating quantum circuit design. This endeavor, known as Quantum Architecture Search (QAS) \cite{martyniuk2024quantum}, integrates diverse search algorithms and machine learning techniques to create optimized quantum architectures tailored for superior performance.Various methods, including reinforcement learning (RL) \cite{kuo2021quantum,ye2021quantum,zhu2023quantum,sogabe2022model,wang2024rnn,fodera2024reinforcement,chen2023quantumRL_QAS,dai2024quantum} and evolutionary algorithms \cite{ding2022evolutionary,chen2024_Evo_QAS_ED}, have been investigated to tackle the complexities inherent in QAS.
Differentiable programming offers an alternative strategy for exploring quantum architectures by parameterizing quantum circuit structures with continuous variables. This approach has been applied within QAS to support various QML tasks, such as optimization, classification \cite{chen2025learning}, and reinforcement learning \cite{zhang2022differentiable,sun2023differentiable,chen2024differentiable}.

A key challenge in both RL-based and differentiable QAS lies in tuning numerous hyperparameters, including exploration-exploitation balances for RL and optimization settings for gradient-based training. In contrast, genetic algorithms do not require gradient information and are inherently well-suited for the combinatorial, non-differentiable circuit spaces of quantum systems, offering robustness to architectural variability and noise. In this paper, we adopt evolutionary optimization for QAS, leveraging its natural parallelizability to evaluate large populations of quantum autoencoders across multiple CPU cores, compute nodes, or QPUs when available.

\subsection{Quantum Autoencoder}
Quantum autoencoders introduce a paradigm shift by leveraging quantum mechanics principles such as superposition and entanglement to perform compression and feature extraction beyond classical capabilities \cite{romero2017}. They have demonstrated potential in quantum state compression, noise reduction, and quantum simulations \cite{bondarenko2019}. For example, by reducing qubit requirements for quantum representations, quantum autoencoders result in more efficient quantum simulations on near-term devices. Additionally, quantum autoencoders have been applied in classical domains, such as image recognition and anomaly detection, by encoding classical data into compact quantum states and reconstructing it with minimal information loss. Our study investigates hybrid quantum-classical autoencoders optimized for image reconstruction, with potential applications in other ML domains.

\subsection{Genetic Algorithms}
Genetic algorithms (GAs) are population-based heuristic search methods inspired by the principles of Darwinian natural selection. They are widely used to find near-optimal solutions in large and complex search spaces by mimicking biological evolution through selection, crossover, and mutation \cite{shrestha2016,hassanat2019}. The process begins with a randomly generated population of candidate solutions, each evaluated by a fitness metric that quantifies performance. Top-performing individuals are selected to pass on their traits to the next generation \cite{raghavjee2015}. Crossover combines portions of selected solutions to create offspring with potentially beneficial traits, while mutation introduces random variations to maintain diversity and prevent premature convergence. This balance of exploration and exploitation allows GAs to improve candidate quality over successive generations \cite{shrestha2016, hassanat2019}.

Compared to traditional optimization techniques—like gradient descent or exhaustive search—GAs are more effective in navigating complex, non-convex, and non-differentiable spaces \cite{carr2014, pratihar2012}. GAs efficiently explore large design spaces \cite{nutini2015, safran2018}, making them well-suited for neural architecture search and quantum circuit design. However, their use in quantum autoencoders is underexplored. Our work addresses the gap by evolving the architecture while using a gradient-based optimizer to update the parameters within hybrid quantum-classical autoencoders.

\section{Quantum Autoencoder}

\subsection{Autoencoders}
An autoencoder~\cite{hinton2006reducing} is a composition of functions that compress data and attempt to reconstruct original data from the compression. Specifically, let a dataset be $\{x^{(j)} \in \mathbb{R}^n \}_{j=1}^N$ of $N$ samples, then an autoencoder consists of two functions $f:\mathbb{R}^n \to \mathbb{R}^m$ and $g:\mathbb{R}^m \to \mathbb{R}^n$ such that the following \textit{reconstruction loss} is minimal,
\begin{equation}\label{E: reconstruction loss}
    L(f, g) = \frac{1}{N} \sum_{j=1}^N \| g(f(x^{(j)})) - x^{(j)} \|^2
\end{equation}
where $f$ and $g$ are called the \emph{encoder} and the \emph{decoder}, respectively. In \cite{hinton2011transforming, liu2004autoencoder, micelucci2022introduction}, autoencoders are constructed by neural networks. The output of the encoder $z^{(j)} := f(x^{(j)}) \in \mathbb{R}^m$ is also called a \emph{latent variable}, where the compression occurs when $m < n$. Then, the reconstruction is performed by the decoder $g(f(x^{(j)}))$ to approximate the original input $x^{(j)}$.

Within the NISQ era, noise plays a critical role in the design and performance of autoencoders, influencing the stability of the training procedure and quality of the reconstructed outputs \cite{cao2021noise, escudero2023assessing}. A substantial amount of noise will cause instability within the quantum autoencoders \cite{mok2024rigorous} due to the degradation of quantum states during computation. Gate errors, decoherence, and measurement noise can lead to deviations in the expected outcomes, corrupting the fidelity of reconstructed data \cite{bondarenko2019}. To stabilize the models, our quantum autoencoders are designed as hybrid quantum-classical autoencoders where the classical sections can help mitigate for the noise-induced errors in the quantum components \cite{liao2024machine}.   
\subsection{Variational Quantum Circuit}
Variational quantum circuits (VQC)~\cite{mitarai2018quantum}, also known as parameterized quantum circuits (PQC), consist of quantum gates with trainable parameters to perform data fitting and ML tasks. A VQC comprises three main components:
\begin{enumerate}
    \item \textit{Feature encoding}: Mapping classical data $x \in \mathbb{R}^n$ into a quantum state via an angle encoding 
    to derive a unitary gate $x \mapsto V(x)$. The resulting state $V(x) \ket{\psi_0}$ by an initial $\ket{\psi_0}$ (or $\ket{0}^{\otimes n} $ in multi-qubits) is then associated with classical data $x$.

    \item \textit{Variational Circuit (Parameterized Quantum Gates)}: Utilize a sequence of tunable quantum gates $\{ U_1(\theta_1), \ldots, U_L(\theta_L)\}$ to form
    \begin{equation}\label{E: unitary sequence}
        U(\Theta) := U_L(\theta_L) U_{L-1}(\theta_{L-1}) \cdots U_1(\theta_1)
    \end{equation}
    where $\Theta = \{\theta_1, \theta_2, \ldots, \theta_L\}$ are trainable parameters to be sought. Each gate $U_\ell(\theta_\ell)$ is typically constructed by $e^{-i \theta_\ell \sigma_k / 2}$ with $\sigma_k$ representing a Pauli matrix.

    \item \textit{Measurement:} After the above two steps, the quantum state is then measured to produce a real-valued prediction using a Hermitian observable $H$ via $\bra{\psi} H \ket{\psi} $.
\end{enumerate}

\begin{figure}
    \centering
    \includegraphics[width=1\linewidth]{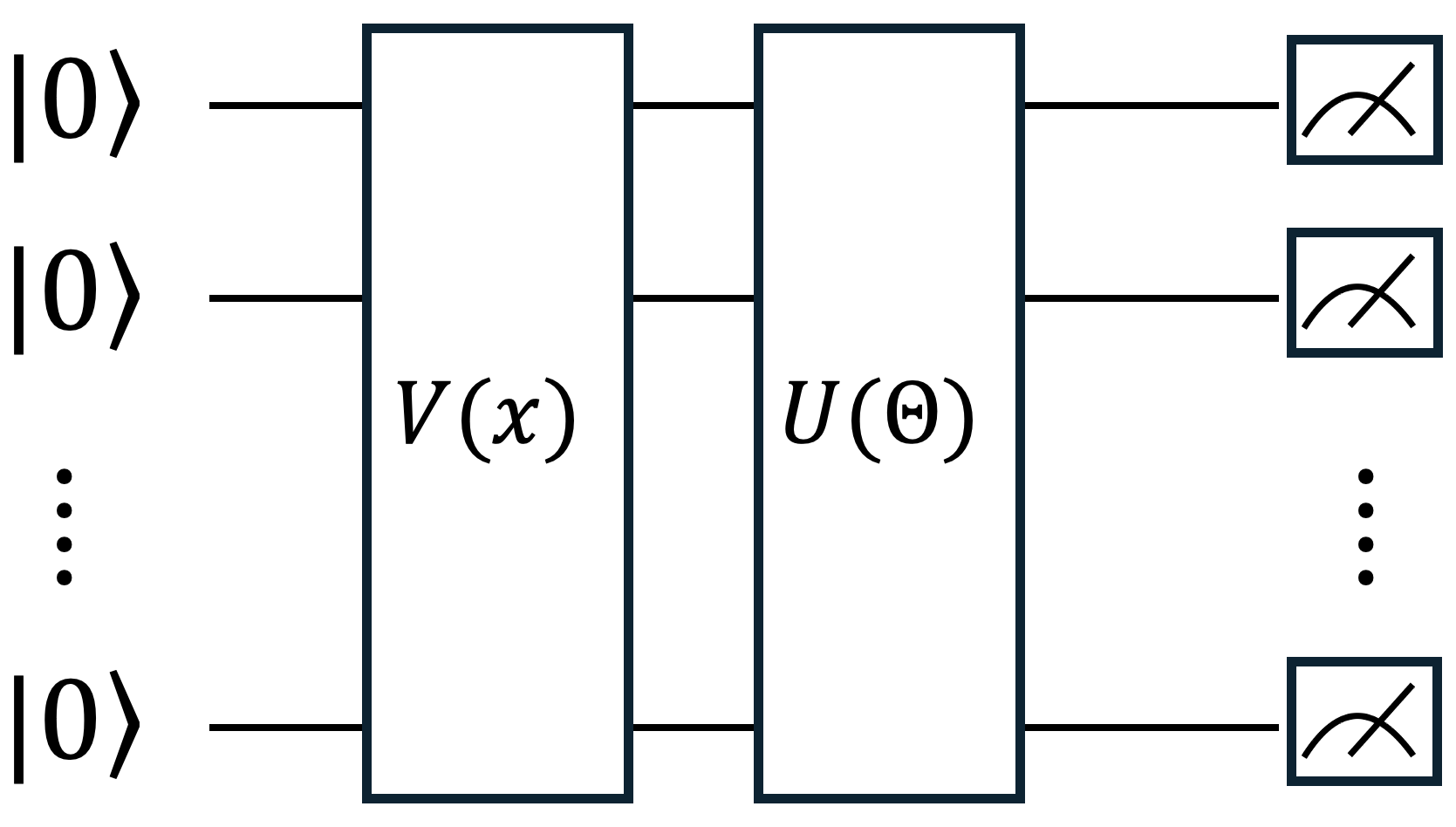}
    \caption{A VQC diagram. A classical input $x$ is encoded using an operator $V(x)$, followed by the application of the variational circuit $U(\Theta)$. The circuit's output is then measured using a predefined observable.}
    \label{VQC Diagram}
\end{figure}

Together, the above three steps give the output of a VQC from an input $x^{(j)} \in \mathbb{R}^n$:
\begin{equation}\label{E: VQC function}
    f_{\text{VQC}}(x^{(j)}; \Theta) = \bra{\psi_0} V^\dagger(x^{(j)}) U^\dagger(\Theta) H U(\Theta) V(x^{(j)}) \ket{\psi_0}
\end{equation}
Then a loss function will be incorporated to search for optimal VQC parameters $\Theta $. \figureautorefname{\ref{VQC Diagram}} provides a visual representation of a VQC, depicting the sequential process of encoding classical inputs, applying trainable quantum gates, and performing measurements. 

This study particularly focuses on optimizing VQCs for improved performance. In the hybrid quantum-classical framework, the VQC can be integrated with other classical (e.g. deep neural networks, tensor networks) models. The hybrid model is capable of performing quantum optimization algorithms, quantum sequence learning \cite{yang2022bert,stein2023applying,chen2024QFWP}, quantum classification \cite{chen2021end,qi2023qtnvqc,lin2024quantumGRADCAM} and quantum reinforcement learning (QRL) algorithms~\cite{chen2022variationalQRL,chen2023quantum}. In our proposed experiment, we are given parameterized gates with a classical optimizer to enhance our model for improved performance.

By virtue of a hybrid quantum-classical model, we construct an autoencoder pair $(f, g)$ by a decoder $g$ of classical network and a hybrid encoder $f = f_{\text{VQC}} \circ f_{\text{C}}$, a classical network $f_{\text{C}}$ followed by a VQC function $f_{\text{VQC}}$ in Eq.~(\ref{E: VQC function}). Mimicking the design of classical autoencoder, the reconstruction loss Eq.~(\ref{E: reconstruction loss}) is also adopted,
\begin{equation}\label{E: VQC reconstruction loss}
    L(f, g) = \sum_{j=1}^N \| (g \circ f_{\text{VQC}} \circ f_{\text{C}})(x^{(j)}) - x^{(j)} \|^2
\end{equation}

The VQC function contains numerous combinations of unitary gates as seen in Eq.~(\ref{E: unitary sequence}). Each specific arrangement of gates yields a unique model with varying performance. This variability suggests the need for a systematic approach to designing quantum circuit architectures. The objective of this study is to search the optimal circuit configurations and gate combinations in $f_{\text{VQC}}$ that minimize the loss defined in Eq.~(\ref{E: VQC reconstruction loss}).

\section{Neural Architecture Search for Quantum Autoencoders}
Our work introduces a novel framework for neural architecture search in quantum autoencoders, leveraging GAs to optimize the structure of VQCs and using standard gradient-based methods for subsequent VQC parameter optimization. Unlike traditional QML approaches that focus solely on parameter tuning, our method systematically explores the design space of quantum circuits, evolving models toward optimal configurations for data reconstruction tasks.

\begin{figure}[!t]
    \centering
    \includegraphics[width=0.45\textwidth]{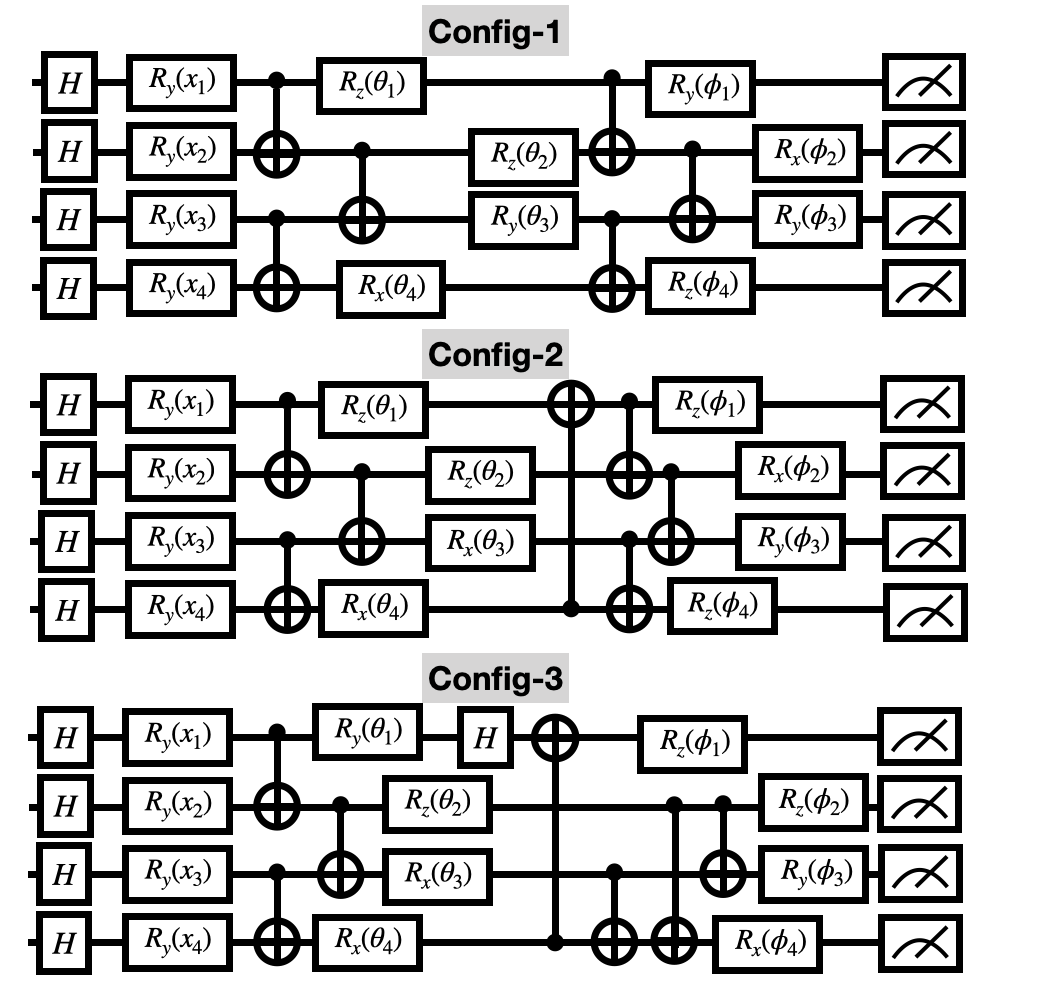}
    \caption{Example of circuit mutations passing down generation by generation.}
    \label{fig:configs}
\end{figure}

We formulate each quantum circuit as an individual in a population, where each ``gene'' represents a quantum gate in the VQC's variational layer. The gate set spans both parameterized (e.g., RX, RY, RZ) and non-parameterized operations (e.g., Hadamard, CNOT), enabling flexible architectural representations. Mutations are applied at the gate level, introducing variations such as gate replacements, parameter adjustments, and the insertion or deletion of non-parameterized gates. As demonstrated in Figure~\ref{fig:configs}, these mutations give rise to increasingly refined and expressive circuit configurations over successive generations.

Each individual in the population encodes a full VQC configuration, where each gene specifies a gate type, its location in the circuit, the target qubit(s), and any associated parameters. This genetic representation supports the global topological mutations (e.g., swapping gate sequences or altering entanglement patterns), while the local changes (e.g., perturbing a rotation angle) are configured by standard gradient methods. Mutation operations are probabilistically applied, and future extensions may include crossover strategies to combine high-performing substructures from multiple parents.

A single quantum autoencoder integrates a fixed feature map to embed classical inputs into quantum states. The variational layer then encodes and entangles these states using trainable quantum gates, which are optimized with classical techniques. Final measurements in the $Z$-basis collapse the quantum information, allowing for reconstruction by classical decoder layers.

Within each generation, models are evaluated using reconstruction loss, and the top three performers are preserved as ``parents''. The remaining population is regenerated through mutation, creating structurally distinct offspring. These mutations not only fine-tune model behavior but also enable exploration of quantum entanglement strategies and circuit depth, broadening the design space well beyond fixed-template models.

As outlined in Algorithm 1, this evolutionary cycle repeats over multiple generations. The goal is to generate a population of quantum autoencoders that converge to low-loss solutions while maintaining structural diversity, which may offer insights into the relationship between circuit design and learning performance.

The core novelty of our approach lies in its ability to evolve both the topology and learning dynamics of quantum models with an autoencoding framework, an area that remains underexplored in QML. This evolution strategy promotes architectural diversity and enhances quantum feature learning, particularly in cases where optimal gate configurations are not known a priori.

\setcounter{algorithm}{0}
\begin{algorithm}[H]
\caption{Neural Architecture Search for Quantum Autoencoders}
\begin{algorithmic}[1]

\State Initialize $population$ with $n\_qubits$ and a set of parameterized gates
\For{$g = 1$ to $generations$}
    \State Train all models (or only mutated ones if $g > 1$) for $epochs$
    \State Evaluate validation loss for all models
    \State Retain top $k$ models with lowest validation loss
    \For{each non-top model}
        \State Replace with mutated copy of a top model
    \EndFor
\EndFor

    \Return Best model and validation loss
\end{algorithmic}
\end{algorithm}

\section{Experiment} 
In this section, we evaluate the performance of hybrid quantum-classical neural networks optimized through a genetic algorithm framework on benchmark datasets. To develop the models, we used PennyLane as our framework. We begin by initializing a population of ten hybrid autoencoders, each configured with fine-tuned parameters. These models are individually trained and assessed based on validation loss. The top three performers are selected to progress to the next generation, while the remaining models are replaced with mutated variants designed to introduce architectural diversity.

Each model follows a consistent architecture: two classical encoder layers, a single VQC layer, and two classical decoder layers. The classical encoders reduce the input dimensionality, after which the compressed data is encoded into quantum states via the VQC’s feature map. Within the VQC, quantum operations entangle the qubits and apply parameterized transformations to extract high-level quantum features. Following the variational ansatz, Pauli-Z measurements convert the quantum state back into classical information, which is then passed through the classical decoders for final reconstruction.

Each VQC is composed of four qubits and a circuit depth of two. This architecture is empirically chosen to balance expressivity and classical simulation cost. Deeper circuits or more qubits demand substantial resources, while shallow designs align with NISQ-era hardware constraints such as gate errors and decoherence. These values were selected to balance circuit expressiveness with computational efficiency: larger or deeper circuits significantly increase classical simulation costs and make it harder to determine whether performance gains stem from improved gate configurations or simply greater capacity. Fixing the number of qubits and circuit depth allows us to isolate the effects of gate-level mutations and topological variation without introducing confounding factors from circuit scale. The quantum layer begins with Hadamard gates followed by RY rotations (serving as the feature map), continues with parameterized gates interleaved with CNOT entangling gates, and concludes with measurement in the Z-basis. We employ a batch size of 256 for optimization, training each model for 10 epochs per generation across five generations in total. This configuration was chosen to enhance convergence stability while maintaining computational efficiency across successive evolutionary cycles.

To evolve circuit performance over generations, mutations are applied specifically to the VQCs, and include modifications such as altering or replacing parameterized gates, as well as adding or removing non-parameterized gates (e.g., Hadamard and CNOT gates) to encourage exploration of entanglement strategies and circuit expressivity. Once mutated, the models are retrained and re-evaluated, iterating this process over multiple generations.

To maintain a balance between exploration and stability, mutation rates for both parameterized and non-parameterized gates are set to 20\%. This value was chosen empirically after observing that it introduced meaningful circuit variations without completely altering the structure of the VQC. This rate ensures sufficient circuit diversity across generations to escape local minima, while avoiding excessive disruption that could destabilize promising configurations. Additionally, the number of parameterized gates initialized in the first generation is kept fixed across generations to prevent architectural drift and ensure training remains focused and consistent, which is especially critical for convergence in quantum neural networks.

\subsection{MNIST}
\begin{figure}
    \centering
    \includegraphics[width=1\linewidth]{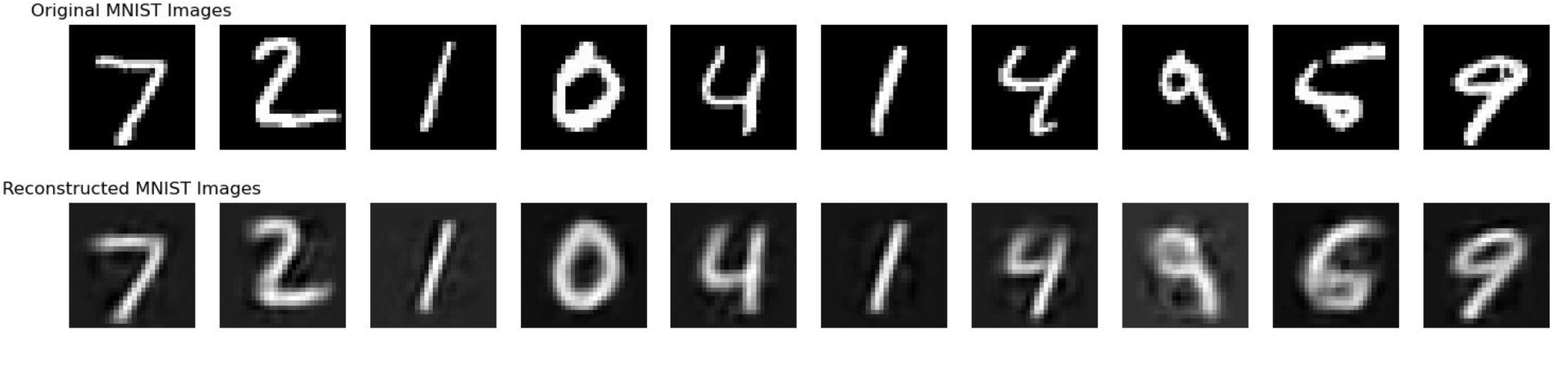}
    \includegraphics[width=1\linewidth]{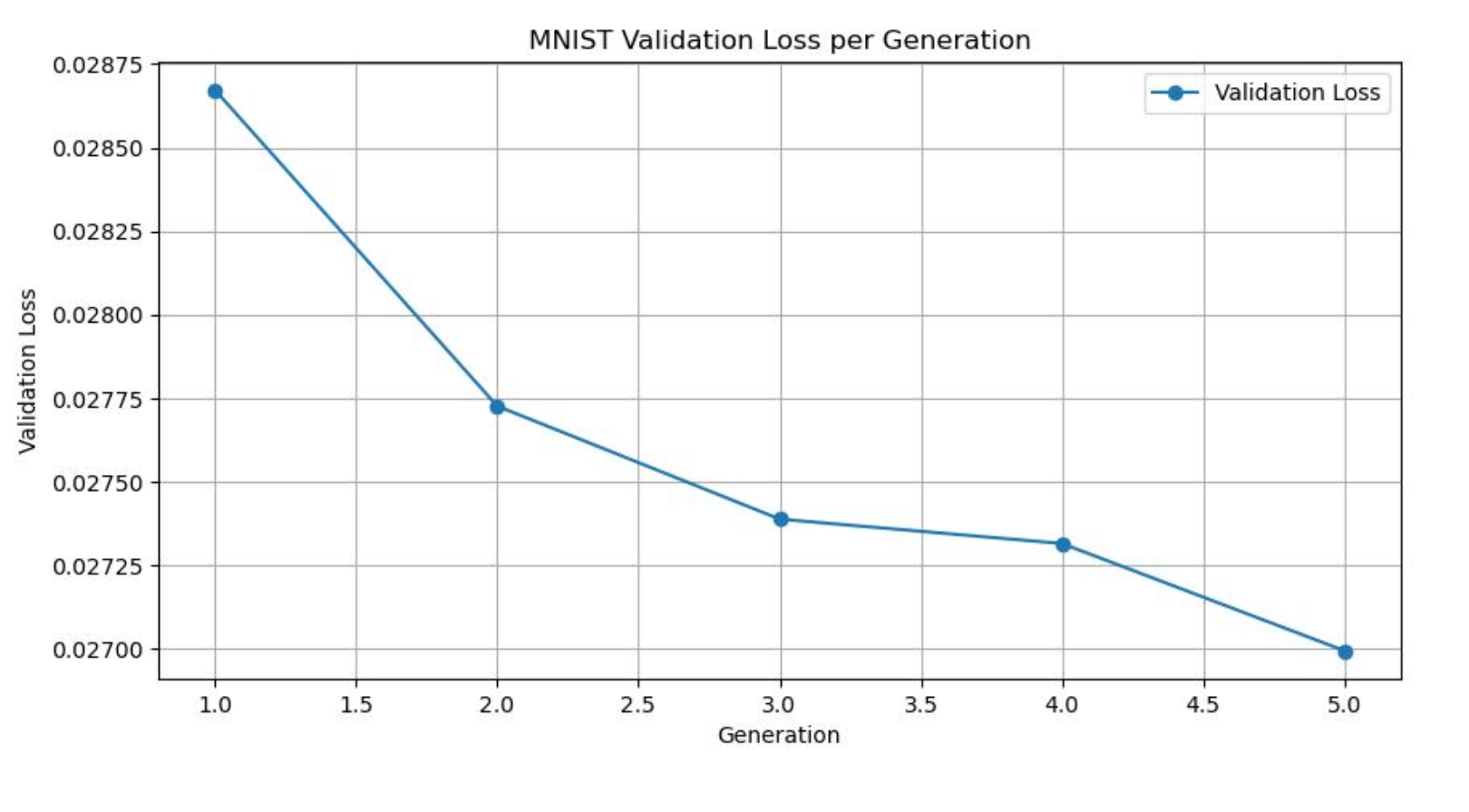}
    \includegraphics[width=1\linewidth]{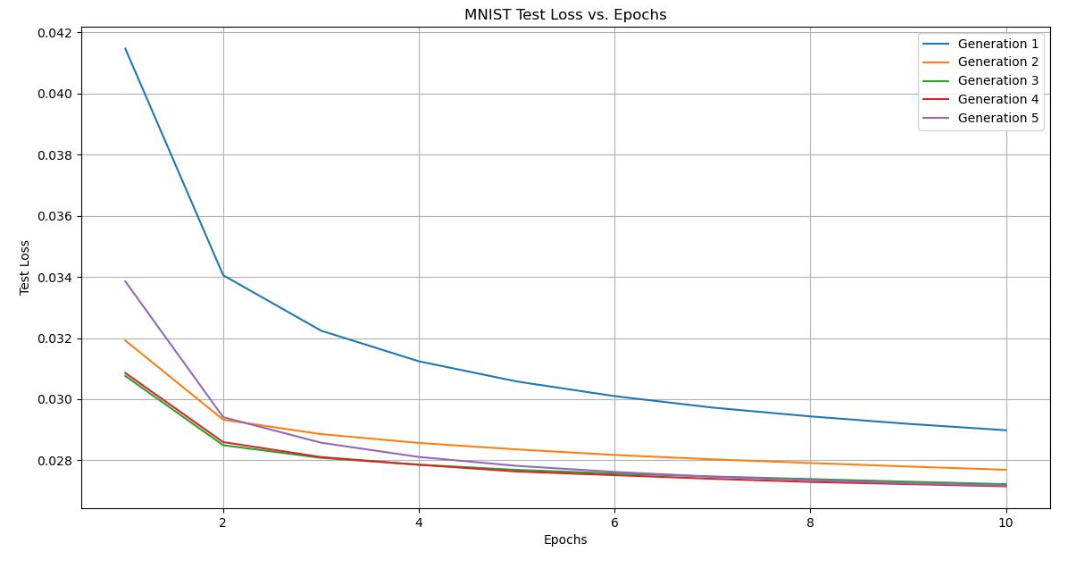}
    \caption{Benchmarking MNIST Data.}
    \label{fig:mnist}

    \centering
    \includegraphics[width=1\linewidth]{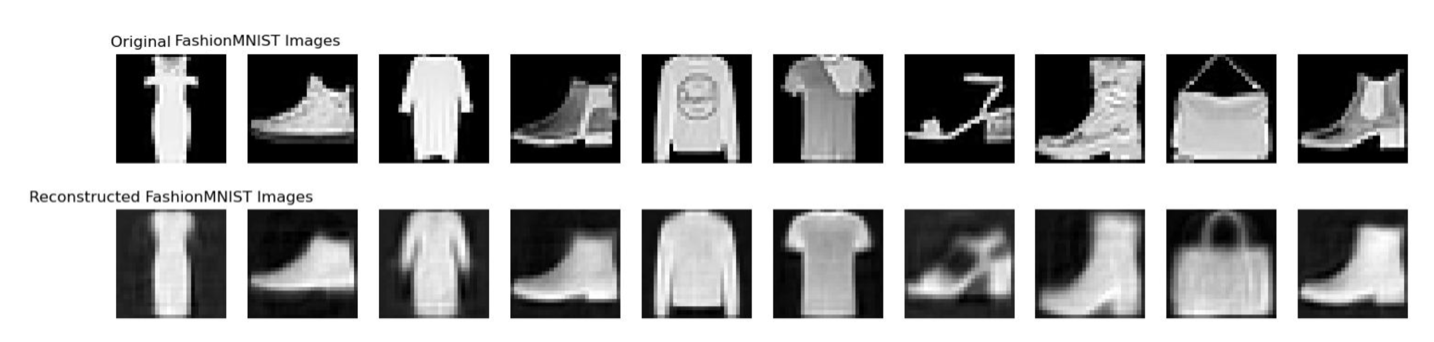}
    \includegraphics[width=1\linewidth]{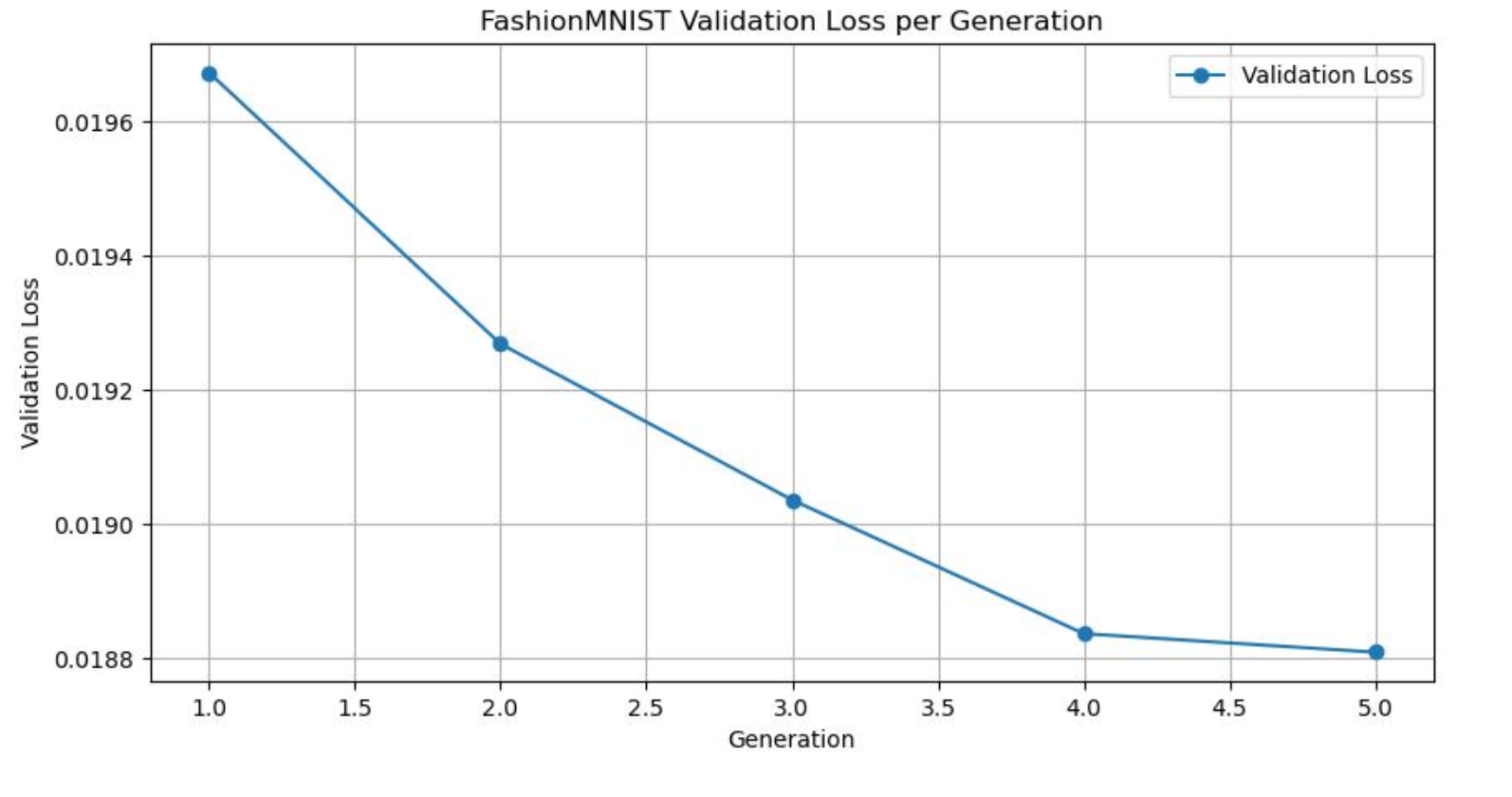}
    \includegraphics[width=1\linewidth]{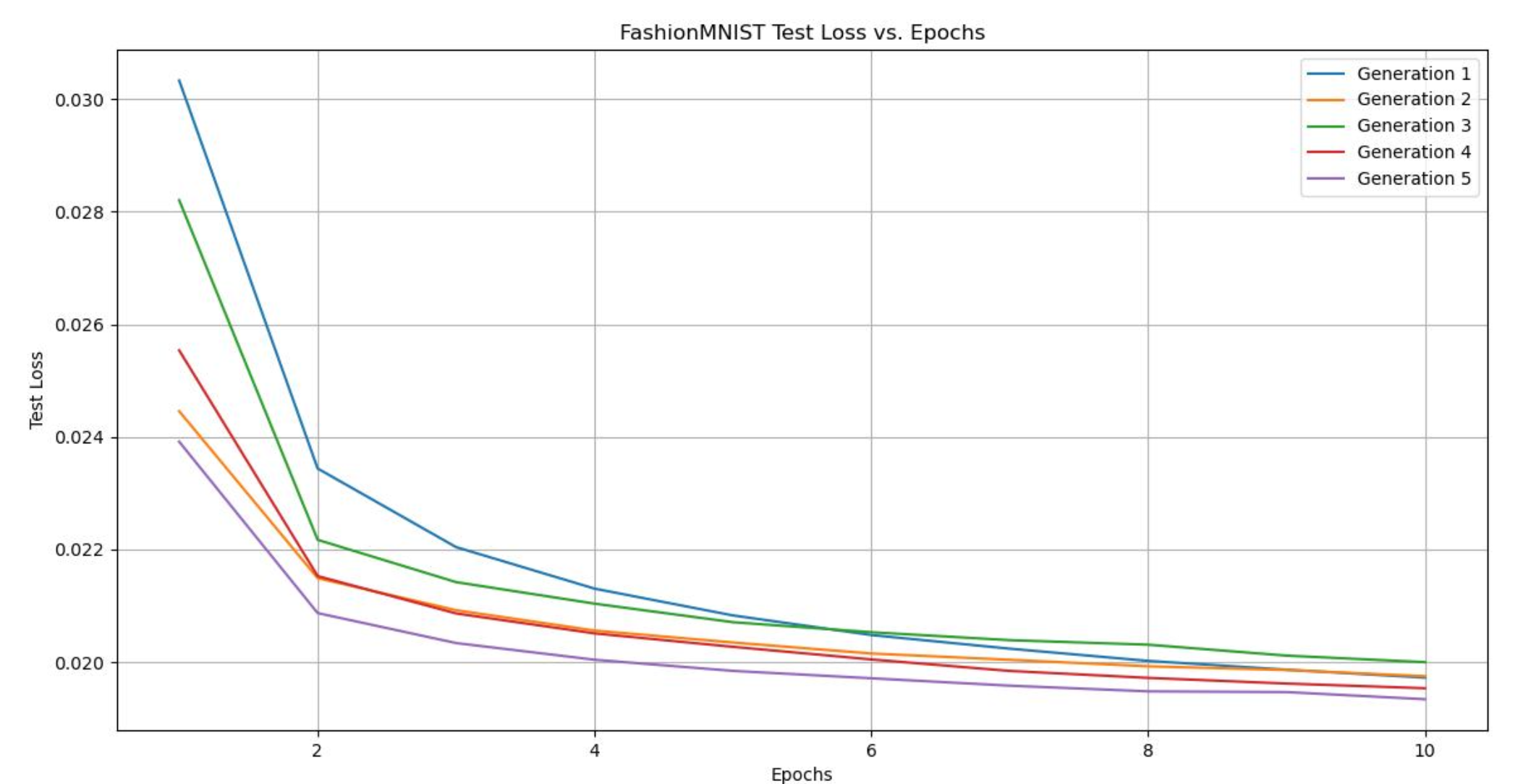}
    \caption{Benchmarking FashionMNIST Data.}
    \label{fig:fashionmnist}
\end{figure}

We perform the reconstruction of the MNIST dataset with our neural network model, which incorporates hybrid quantum-classical layers for dimensionality reduction and data reconstruction. We trained the model with a batch size of 256, optimizing performance through an evolutionary process. Each experiment begins with a population of 10 autoencoders trained on the MNIST dataset. After evaluation in the first generation, the three top-performing models are preserved unchanged, and the remaining seven models in the next generation are formed by mutating these top performers. In subsequent generations, only the mutated models are trained, while the top three from the previous generation are retained without further modification. The final performance of our trained model is illustrated in \figureautorefname{\ref{fig:mnist}}, which presents the results of the full training process. The top portion of the figure shows an original MNIST image alongside its reconstructed counterpart, demonstrating the model's ability to capture key image features. The middle plot depicts the validation loss trend across generations, indicating steady improvements in reconstruction accuracy. The bottom graph benchmarks model performance by showing test losses averaged across all trained agents during each generation, demonstrating the generalization performance of the evolving population of autoencoders. These results validate the effectiveness of our genetic algorithm in refining quantum autoencoder architectures. 

\subsection{FashionMNIST}
We extend our experiment to the FashionMNIST dataset using the same neural network architecture and training parameters as in the MNIST case. Similarly, the initial population consists of 10 trained autoencoders. In each subsequent generation, the three fittest models are preserved, while the remaining population is filled by mutated offspring of these top performers. This experiment assesses the generalizability of our quantum-classical autoencoder across different image types. \figureautorefname{\ref{fig:fashionmnist}} presents the reconstructed images alongside their originals, illustrating the model's ability to capture key features such as clothing textures and structural details. The middle graph shows a consistent decline in validation loss per generation as the genetic algorithm optimizes circuit configurations. The bottom graph highlights a reduction in test loss across epochs, reflecting averages computed from all agents trained during each generation on the FashionMNIST dataset. These results confirm the effectiveness of our approach in learning meaningful representations across diverse image datasets.

\subsection{Results}

Our results show that mutated circuits generally outperform unchanged ones. In particular, mutations that add extra entanglements, especially between non-adjacent qubits, tend to lower the reconstruction loss. However, we find that adding too many entangling gates, especially redundant CNOTs, can impair performance. These overly entangled circuits tend to increase reconstruction loss slightly, likely due to unnecessary complexity in the quantum state evolution which can hinder effective training. For example, in cases where the baseline loss was 0.02721, circuits with excessive entanglement exhibited losses around 0.02937 which is a modest increase of approximately 8\%. In general, the loss increase from over-entanglement tends to range between 5–10\%. That said, this trend is not absolute: certain overly entangled circuits, when carefully structured, outperformed their baselines, suggesting that entanglement is beneficial when introduced judiciously. Modifications to parameterized gates and additions of non-parameterized gates may improve or slightly degrade performance, with some mutations surpassing the original circuits' loss and others producing only minor fluctuations in loss.

\section{Conclusion}
We introduced a neural architecture search algorithm that evolves the ansatz of our hybrid quantum-classical autoencoders, enabling joint optimization of both circuit structure and trainable parameters for effective data reconstruction. 

Our results demonstrate that quantum autoencoders, when integrated into a hybrid architecture, can enhance feature learning while reducing computational overhead. This work highlights the potential of evolutionary search as a powerful tool for designing expressive and efficient quantum models. Beyond improving reconstruction performance, our approach promotes architectural diversity and paves the way for principled exploration of quantum circuit design.

Looking ahead, we aim to extend this framework to larger-scale datasets, incorporate adaptive mutation strategies, and investigate the generalization capabilities of evolved quantum architectures across broader machine learning tasks and hardware constraints.

\footnotesize
\bibliographystyle{ieeetr}
\bibliography{refs,bib/qas,bib/qml_examples,bib/autoencoder,bib/ML,bib/vqc,bib/tools,bib/quantum_autoencoder,bib/genetic_algorithms,bib/qc_general}

\end{document}